# Stress balance in nano-patterned N/Cu(001) surfaces

S. Hong,[1] T. S. Rahman,[1*] E. Z. Ciftlikli,[2] and B. J. Hinch[2]

[1]Department of Physics, University of Central Florida, Orlando, FL32816
[2]Dept. of Chemistry and Chemical Biology, Rutgers University, Piscataway, NJ 08854

PACS numbers: 68.43.-h, 62.40.+i

**Abstract**

We employ helium atom scattering (HAS) and density functional theory (DFT) based on the ultrasoft pseudopotential scheme and the plane-wave basis set to investigate the strain and stress balance in nano-patterned N/Cu(001) surfaces. HAS shows that, with increasing N coverage (and decreasing stripe widths), the stress-relief-driven lateral expansion of the averaged lattice parameter within finite-sized N-containing patches reduces from 3.5% to 1.8% and that, beyond a critical exposure, the lateral expansion of the patches increases again slightly, to 2.4%. The latter implies that in this higher coverage range the compressive stress is partially relieved via another mechanism, which turns out to be nucleation of Cu-vacancy trenches. In full agreement with the above and previous experimental observations, DFT calculations show that an optimized N-induced *c(2×2)* structure has a net surface stress level of 4.2 N/m and such stress is effectively relieved when stripes of clean Cu(001) form along the ⟨100⟩ direction or when trench-like steps of Cu atoms form along the ⟨110⟩ direction. Additionally, the calculations demonstrate that (contrary to earlier suggestions) rumpling displacements within the outermost Cu layer do not act to relieve the compressive surface stress levels and that, while clock-like displacements could relieve stress levels, such displacements are energetically unstable.

## I. INTRODUCTION

For several decades experimental and theoretical studies have provided a great deal of insight into the nature of the bonding among atoms and molecules chemisorbed on surfaces and those of the surface. Chemisorption of N on Cu(001) has been the subject of many investigations because the Cu(001) surface displays a striking long-range nanoscale ordering in the course of N adsorption. Nano-sized clusters are nearly square, consisting of N patches with a dimension of about 55Å ×55Å, which locally exhibit *c(2×2)* symmetry, and surrounding clean-surface stripes[1], as indicated in Fig. 1a. The arrangement of N patches across stripes can be either in-phase or out-of-phase, depending on the thickness of those stripes.[2] As N coverage proceeds to saturation (and with probable missing Cu row formations in the N patch boundaries[3]) the island evolves into a nearly homogeneous distribution of N atoms on Cu(001) – N atoms occupying every alternate hollow adsorption site – and the Cu-vacancy trenches forming along the ⟨110⟩ direction, as shown in Fig. 1b.[4] The remarkable nano-patterning is generally believed to be driven by stress relief of the N-rich overlayer.[5]

FIG. 1. (Color online) Schematic diagrams of reconstruction models: **(a)** N patches with stripes **(b)** N patches with trenches **(c)** clock displacement, and **(d)** rumpling displacement. (N: red circle; Cu: white (1st layer), gray (2nd layer), black (3rd layer) circles; Arrow: lateral displacements of Cu atoms in the 1st layer; +/−: vertical displacements of Cu atoms in the 1st layer.)

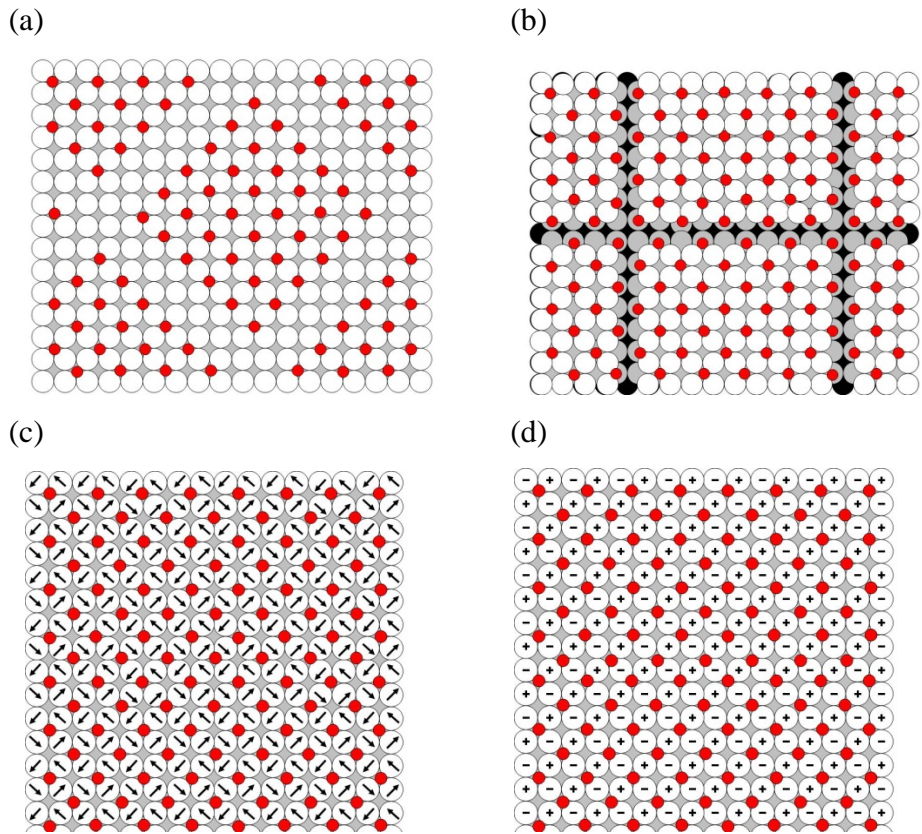

Several models have been proposed with significantly modified substrate structure beneath the patches, the most prominent of which are the clock reconstruction model[6] and the rumpling model.[7,8] These models differ from the prevailing picture of $c(2\times2)$-N Cu(001) surface, which does not assume any reconstruction or significant distortions of Cu at the surface. In the clock model, Cu atoms in the outermost layer shift clockwise or counter-clockwise while N atoms maintain a $c(2\times2)$ site registry, as in Fig. 1c. The rumpling model postulates a large rumpling of 0.34 Å in the outermost Cu layer and a commensuration of N patches with the substrate as in Fig. 1d. The rumpling model is based on the data from Photoelectron Diffraction (PhD) measurements[7] and Scanning Tunneling Microscopy (STM) experiments,[8] in which bright spots in the images were interpreted as Cu rather than N atoms. However, neither the clock nor the rumpled model has so far been supported by subsequent experiments or theoretical calculations. More recent STM experiments,[9-12] for example, have not indicated such reconstructions at any N coverage. Instead the majority of these experiments interpret the bright spots as N atoms being incommensurate with the substrate.[9-11] DFT-GGA calculations[13,14] also show that for various striped structures (with different stripe-to-stripe distances) N-atom separations are not commensurate with the substrate Cu atoms, that rumpling in inner Cu atoms is less than 0.15 Å, and that N atoms always sit approximately above the first-layer Cu atoms. The STM images simulated in these studies also show that N atoms appear bright. Briefly, these more recent experimental and theoretical studies consistently predict unrumpled, unreconstructed substrates as the basis for $c(2\times2)$ N/Cu(001) surfaces.

The interest in the above two reconstructed models lies in the fact that they propose novel surface-stress relief mechanisms. In both, stress relief is accompanied by elongation of Cu-Cu bonds in the top layer. In the rumpling model, this is caused by a large vertical displacement (rumpling) in the top Cu layer. In the clock model, it results from lateral displacement (rotation) in that top Cu layer, in analogy to what happens in the C/Ni(001) system.

Since each model purports to bring about stress relief, calculations of surface stress can directly address those assumptions. Although the experimental and calculated surface stresses have already been reported for clean and $c(2\times2)$-N/Cu(001) surfaces,[14-16] surface stress calculations of the alternative structures, implied by the competing models, have not been performed. We have thus carried out first-principles density-functional calculations to evaluate surface stress levels in ideal *c(2×2)*-N/Cu(001), in surfaces with experimentally-observed N-free stripes of various boundary geometries and thicknesses, in surfaces with experimentally-observed Cu-free trenches of various directions and thicknesses, and in other hypothetical surfaces such as rumpled and clock-reconstructed Cu(001). We demonstrate the effectiveness of the inter-island boundaries for stress relief in *c(2×2)*-N/Cu(001) surfaces by deriving the one-dimensional (1D) stress formula as the function of the periodicity of stripes and stripe width.

We also used helium atom diffraction to investigate the growth of half-order (1/2, 1/2) diffraction features in annealed surfaces with increasing N exposures, observed at room temperature, in order to examine possible domain ordering and observed surface strain changes in the N-containing domains. The N-N spacing, of course, must reflect surface Cu lattice parameters and hence also surface stress levels. We conclude that the stress levels can increase as the N coverage initially increases, but that at the high N coverages the surfaces (with coexisting stripes and trenches) show decreasing stress levels with increasing N-coverage. We set out details of our theoretical and experimental methods in Sec. II, present our results and discussions in Sec. III, and summarize our conclusions in Sec. IV.

## II. METHODS

### A. Theoretical methods

We based our DFT calculations on the plane wave basis set[17] and the ultrasoft pseudopotential scheme.[18] We used the Quantum-Espresso computer code.[19,20] For the exchange-correlation energy, we have used the generalized gradient approximation (GGA) with the Perdew-Burke-Ernzerhof functional.[21] We set the kinetic energy cutoff at 544 eV for the plane-wave basis set. The calculated lattice constant for Cu bulk was 3.67Å, which is 1.6% larger than that of experiment.

We use several surface models to study structural relaxations and surface stresses in *c(2×2)*-N/Cu(001) surfaces. For simulating an ideal *c(2×2)*-N/Cu(001) surface, we used the *c(2×2)* unit cell. We also used *c(2×2)* for a rumpled Cu(001) structure, and necessarily a *p(2×2)* unit cell for the clock reconstruction. Figures 2 to 5 show schematic diagrams of the larger unit cell structures with N-free stripes or Cu-free trenches that are studied here. To mimic N patches, we introduce into our surface model a 1D stripe aligned along the ⟨100⟩ direction with a N-patch of width $l = 1, 3, 4\ a_o$ ($a_o$: lattice constant, 3.67Å) and stripe widths $d = 1, 2\ a_o$, such that the corresponding surface unit cells are *(2√2×√2)R45°, (4√2×√2)R45°* and *(5√2×√2)R45°*, as in Fig. 2a-d. These striped surface models assume in-phase boundaries. In order to study the effect of out-of-phase boundaries we used the surface unit cell of $\begin{pmatrix} 5 & 4 \\ 0 & 1 \end{pmatrix}$ with a N-patch of widths $l = 3, 4\ a_o$ and stripe widths $d = 1.5, 0.5\ a_o$ as in Fig. 3a and 3b. To study the formation of a missing Cu row in the monoatomic-wide stripe boundary, we used *(5√2×√2)R45°* unit cell with a N-patch of width $l = 4\ a_o$ and stripe width $d = 1\ a_o$ as in Fig. 4. For trench formation, we align a 1D trench either along the ⟨110⟩ direction, as in Fig. 5a-c, or along the ⟨100⟩ direction, as in Fig. 5d-e, to examine the effect of the alignment direction, with the N-patch width $l = 1/\sqrt{2}, 1, 3/\sqrt{2}, 5/\sqrt{2}, 4\ a_o$ and trench width $d = 1, 1/\sqrt{2}\ a_o$, such that the corresponding surface unit cells are p(2×2), *(2√2×√2)R45°*, p(4×2), p(6×2), and *(5√2×√2)R45°*, as indicated in Fig 5. We then calculated surface stresses in the above *c(2×2)*-N/Cu(001) surfaces in the surface direction perpendicular to stripes or trenches. With the calculated lattice constant $a_o = 3.67$Å, N-patch width in the above surfaces varies from 11 to 15 Å.

The supercell consisted of slabs of nine Cu layers for *c(2×2), p(2×2)*, and *(2√2×√2)R45°* unit cells, and of four Cu layers for *(4√2×√2)R45°, (5√2×√2)R45°*, $\begin{pmatrix} 5 & 4 \\ 0 & 1 \end{pmatrix}$, *p(4×2)*, and *p(6×2)* unit cells. On each side of the nine-layer slabs, N overlayers were adsorbed symmetrically with respect to the center layer, and all atoms were allowed to relax maintaining inversion symmetry. In the four-layer slab calculations, however, an N overlayer was placed only on one surface and two Cu layers on the clean side were fixed to the bulk positions, and all (free) atoms were allowed to relax until the forces on them fell below $2\times10^{-2}$ eV/Å. A schematic diagram with labels relevant to structural parameters is shown in Fig. 6, where dN-N and dCu$_1$-Cu$_1$ are the (lateral) nearest neighbor N-N and Cu-Cu distances in the topmost layer, respectively, $d_{N-Cu_i}$ and $d_{ij}$ are the vertical interlayer distances between N and Cu atoms in the *i-th* Cu layers (from the top) and between Cu-Cu in the *i* and *j-th* Cu layers, respectively, and $r_i$ and $\delta$ are rumpling in the *i-th* Cu layers and lateral shift of Cu atoms in the topmost layer, respectively. For statistics, we evaluate these structural parameters only for atoms within N patches (thus not including Cu atoms in stripes) except for rumpling $r_i$, which is averaged for all Cu atoms in the *i-th* Cu layer.

The vacuum space between the supercell and its periodic images was in excess of 9 Å. We employed the Monkhorst-Pack scheme[22] for the following k-point sampling of the Brillouin zone: (9×9×1), (6×6×1), (3×6×1), (2×8×2), (2×10×2), (3×6×2), and (2×6×2) grids for *c(2×2), p(2×2), (2√2×√2)R45°, (4√2×√2)R45°, (5√2×√2)R45°* and $\begin{pmatrix} 5 & 4 \\ 0 & 1 \end{pmatrix}$, *p(4×2)*, and *(6×2)* unit cells, respectively, with a Fermi level smearing[23] of 0.27 eV.

FIG. 2. (Color online) Schematic drawings for N patches with stripes in the in-phase N arrangement with different N-patch width *l* and stripe width *d*: **(a)** $l = 1\ a_o$, $d = 1\ a_o$; **(b)** $l = 3\ a_o$, $d = 1\ a_o$; **(c)** $l = 3\ a_o$, $d = 2\ a_o$; **(d)** $l = 4\ a_o$, $d = 1\ a_o$.

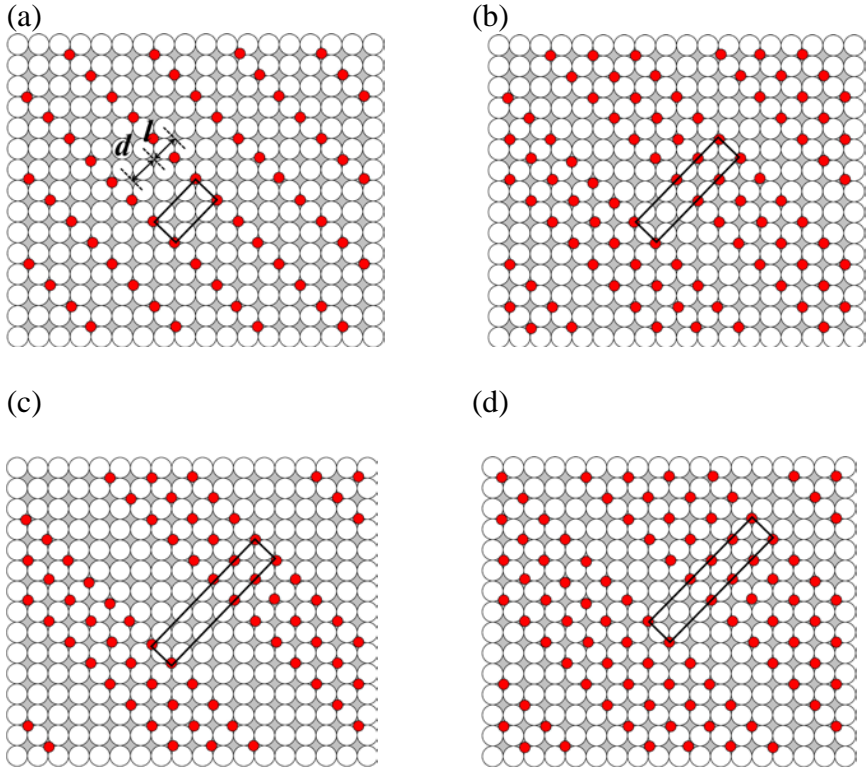

FIG. 3. (Color online) Schematic drawings for N patches with stripes in the out-of-phase N arrangement with different N-patch width *l* and stripe width *d*: **(a)** $l = 3\ a_o$, $d = 1.5\ a_o$; **(b)** $l = 4\ a_o$, $d = 0.5\ a_o$;

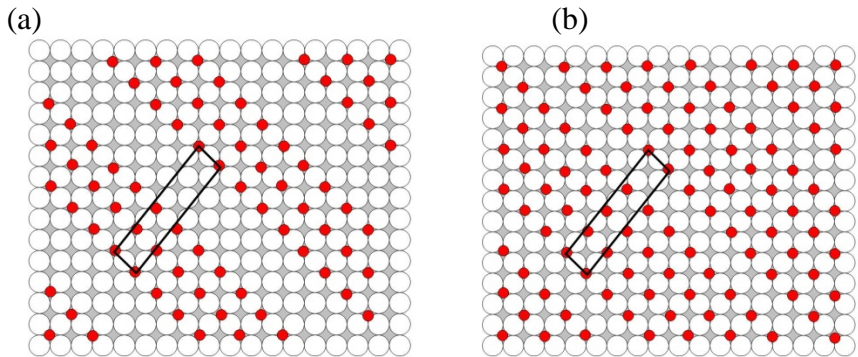

FIG. 4. (Color online) Schematic drawing for N patches with the missing-Cu row boundary ($l = 4\ a_o$, $d = 1\ a_o$).

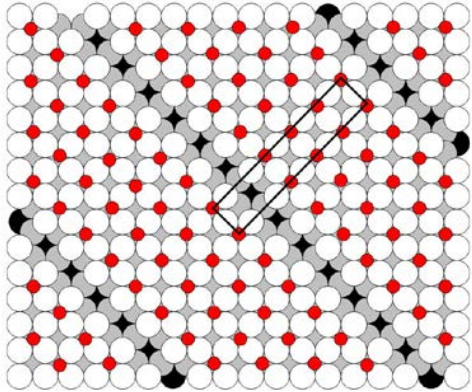

FIG. 5. (Color online) Schematic drawings for c(2×2) N/Cu(001) surfaces with different N-patch width $l$ and trench width $d$. For trenches along <110> direction: **(a)** $l = 1/\sqrt{2}\ a_o$, $d = 1/\sqrt{2}\ a_o$; **(b)** $l = 3/\sqrt{2}\ a_o$, $d = 1/\sqrt{2}\ a_o$; **(c)** $l = 5/\sqrt{2}\ a_o$, $d = 1/\sqrt{2}\ a_o$. For trenches along <100> direction: **(d)** $l = 1\ a_o$, $d = 1\ a_o$; **(e)** $l = 4\ a_o$, $d = 1\ a_o$.

(a)
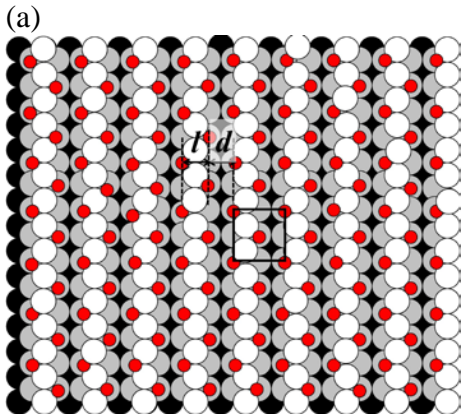

(b)
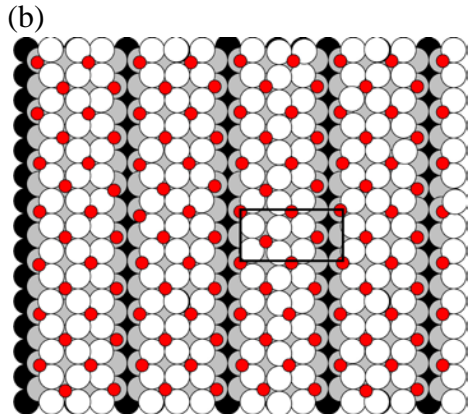

(c)
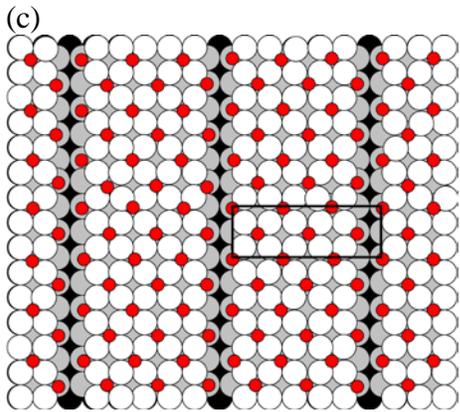

(d)
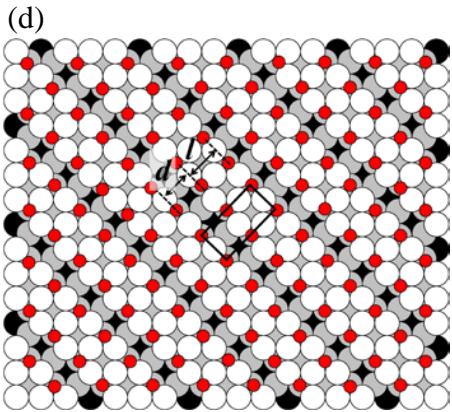

(e)

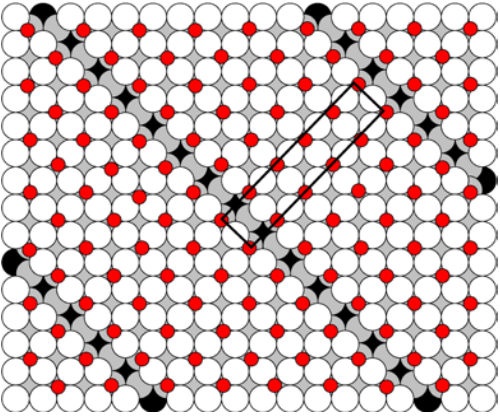

FIG. 6. (Color online) Structural parameters of interest.

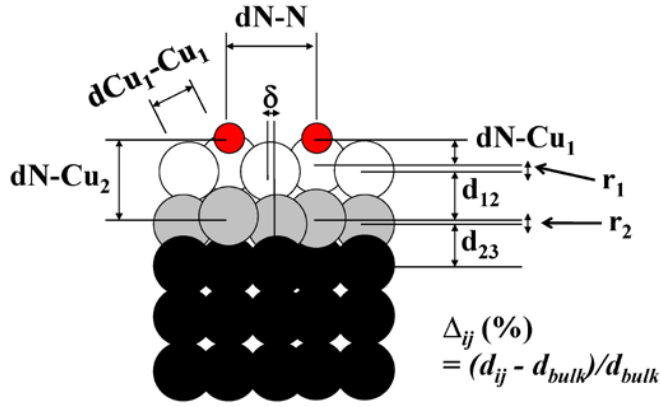

To extract surface stress via *ab initio* methods, we used not only an analytical method but also the standard numerical method in order to avoid systematic errors in the calculations. While the numerical method makes use of calculated derivatives of surface energy with respect to small applied strains, the analytical method uses the stress theorem[24] and requires appropriate corrections to the fictitious stress components that arise from the finite size of the plane-wave basis set. For numerical stress, the strains $\varepsilon$ of ±2% and ±4% (in some cases, ±1% and ±3%, too) are applied and only the diagonal stress components ($\sigma_x$ and $\sigma_y$) were calculated using the following equations:

$$E^{surf} = E^{slab} - E^{bulk} \frac{N^{slab}}{N^{bulk}} \qquad (1)$$

and

$$\sigma_i^{surf} = \frac{1}{A}\frac{dE^{surf}}{d\varepsilon_i}, \qquad (2)$$

where $E^{surf}$, $E^{slab}$, $E^{bulk}$, and $A$ are surface energy, slab and bulk energy, and surface area, respectively. To extract the surface stress $\sigma^{surf}$ from the analytically-calculated slab and bulk stresses, $\sigma^{slab}$ and $\sigma^{bulk}$, we used

$$\sigma_i^{surf} = t^{cell}\left[\sigma_i^{slab} - \sigma_i^{bulk}\frac{N^{slab}}{N^{bulk}}\frac{V^{bulk}}{V^{cell}}\right] \qquad (3)$$

where $N^{slab}$ is the number of Cu atoms in the slab (which is used to calculate $\sigma^{slab}$), and $N^{bulk}$ is the number of Cu atoms in the volume of bulk unit cell, $V^{bulk}$, (which is used to calculate $\sigma^{bulk}$), and $V^{cell}$ and $t^{cell}$ are the volume and thickness of the supercell, respectively, which includes the slab and the vacuum. For the symmetric nine-layer slab, $\sigma_i^{surf}$ in equation (3) is divided by a factor of 2 to account for two identical surfaces. For the asymmetric four-layer slab, the surface stress of the N-adsorbed surface was extracted by setting the stress of its clean bulk-terminated (bottom) surface to that of a bare bulk-terminated four-layer slab.

We should point out that even in fully-relaxed slabs subject to the force threshold ($2\times10^{-2}$ eV/Å), we find non-zero residual stress ($\sigma_z$) along the surface normal, the magnitude of which is mostly, but not always, an order of magnitude smaller than the horizontal stress components ($\sigma_x$ and $\sigma_y$). Furthermore, although in some calculations, the values of $\sigma_{i\,(i=x,y,z)}$ may fluctuate from one ionic iteration to another, the difference $\sigma_{i\,(i=x,y)} - \sigma_z$ always converges. We used this fact to extract reliable stress values from the fluctuating stress components in fully-relaxed slabs, enabling us to set $\sigma_i^{slab}$ in Equation (3) to $\sigma_{i\,(i=x,y,z)} - \sigma_z$. This guarantees that the stress along the surface normal is zero, as it should in fully-relaxed slabs. We also find that it is important to use identical supercells for both slab and bulk calculations in Equation (3). Stress values so calculated are in good agreement with numerical stress values in most cases studied here.

### B. Experimental methods

The high-resolution helium atom scattering apparatus is fully described elsewhere.[25] An intense nearly monoenergetic ($\Delta v/v \sim 1\%$) thermal energy He beam is scattered from the sample crystal, and diffracted He atoms are mass selected and detected in a pulse-counting RF quadrupole mass-spectrometer. Time-of-flight energy analysis confirmed that the incident energy, for all the measurements reported here, was fixed and stable at 31.3 meV. The time-of-flight path length of this instrument has been calibrated using seeded HD in He beams such that $J = 0 \rightarrow 1$ HD rotational energy losses/gains agree with known gas phase values. The fixed scattering angle, 99.0°, is also known to within 0.1°; first-order diffraction peak positions are accurately predicted to well within 0.5%, or typically ~ 0.01 Å$^{-1}$.[25]

In a base pressure of below $2 \times 10^{-10}$ mbar, the single-crystal 1cm-diameter Cu(001) sample was cleaned with repeated cycles of sputtering (15 min, Ne$^+$, 1 keV, ~13 μA/cm$^2$) at RT and annealing at 675 K for 10 min. The same ion gun was used for 1 keV N$^+$ and/or N$_2^+$ ion exposures.

For each nitrogen ion exposure the total current was monitored on the room temperature grounded sample (typically of order ~ 5 μA/cm$^2$). The time-integrated current is reported as a measure of the cumulative N implantation on/into the surface. Any single charge exposure of larger than 3000 μC is known to produce an N-saturated surface, with 0.49 ML of N (after annealing).[5] (Doubling the N exposure produced no discernable further increase in the N Auger signals.)

For the measurements presented here, smaller N-ion doses were made. Following each N-ion implantation the sample was annealed to 600 - 620 K for 5 minutes before each helium-diffraction scan was made. Successive N doses followed previous implantations/anneals. Here the absolute N coverages could not be measured and are unknown, although increasing exposures will increase successive total N coverage in the range between zero and slightly below 0.5ML. Presented here are the results for one series of exposures and anneals. Exactly the same trends were observed with other sequences."

## III. RESULTS AND DISCUSSIONS

### A. HAS experiment

In our He-diffraction measurements, we have not been able to measure the stripe periodicity within the specular, (0,0), feature. The implication is that the apparent average height of the surface, as seen by incoming He atoms, is insensitive to the local nitrogen coverage. Importantly however, half-order intensities were seen to grow even at exposures as low as 50 μC, as illustrated in Fig. 7a. This is fully consistent with isolated domain growth, even at coverages below 0.075 ML. No evidence for domain ordering (peak splitting) was seen in the half-order intensities, at any coverage. This confirms that the relative phases of adjacent *c(2×2)* domains are initially, and remain, uncorrelated at all exposures we investigated. In helium scattering, the main order diffraction peaks remain well below $10^{-2}$ of the specular peak intensities. These peaks may show some signs of peak splitting, at higher N coverages but, given the signal-to-noise levels in the available data, extensive investigations of these peaks were not warranted.

He-atom diffraction has a distinction from other (more conventional) diffraction techniques, which we have exploited in this investigation of surface lattice parameters. The difference between techniques lies in the extreme surface sensitivity of He diffraction, and its reliance on a surface corrugation profile to give rise to finite diffraction intensities. We shall see that, in contrast to the X-ray diffraction or Low Energy Electron Diffraction (LEED) techniques, non-specular He-atom diffraction intensities are dominated by top-most structure within the more corrugated N-containing regions alone. Hence an incommensuration/net expansion of these regions is reflected in the mean half-order peak positions.

An adsorbate-free Cu(001) exhibits a He atom scattering surface corrugation that is typically believed to be $< 10^{-2}$ Å. The unit cell of an uncorrugated (flat) surface therefore does not have a significant form factor at wide scattering angles. Only as the corrugation increases does the form factor distribution spread to wide scattering angles, and its magnitude can become measurable at the positions of half order He diffraction peaks. Thus only the N-containing regions contribute scattering amplitudes that are significant (measurable) around the half-order scattering positions. In our studies the N-containing regions exhibit no discernible phase correlations (and resultant narrowing/splitting of the diffraction peaks). We see therefore an intensity distribution that reflects the magnitude squared of the form factor of single (isolated and laterally extended) N domains. That form factor is peaked at the incommensurate half-order position, $(\pi/a´, \pi/a´)$ where $a´$ is a lattice parameter averaged over only the extent of an N-containing region. The substrate-defined half-order position is at $G/2 = (\pi/a, \pi/a)$. For the uncorrelated expanded N patches, the He intensity is thus peaked at $G/2 – \delta´ (\pi/a, \pi/a)$, where $\delta´ = (a´ – a)/a´$.

In X-ray diffraction, the probe is comparatively insensitive to the N adsorbate. Much stronger diffraction is seen from the bulk and near-surface Cu centers. The half-order peaks observed in X-ray diffraction are therefore dominated by the local N-induced modulation of Cu core positions that are strongest in subsurface layers. The weak X-ray "(½, ½)" diffraction peak intensities may thus occur more closely to the exact G/2 positions. So far as we are aware, their reciprocal space positions have as yet not been accurately analyzed.

Low energy electron diffraction (LEED) is more sensitive to the N adsorbate than are X-rays, but is still sensitive as well to subsurface Cu atomic centers. The half-order peaks observed in LEED are therefore also influenced strongly by the N-induced modulation of subsurface Cu core positions. An isolated N domain is expected to show a net expansion in the topmost Cu layer, but lower layers (exhibiting a 2x2 rumpling periodicity) are not expected to be so strongly laterally expanded. In addition to this depth characteristic of the "half order periodicity," this periodicity is expected to be manifest also in a selvedge region between the N domain and surrounding regions. This selvedge (N-free region) around isolated domains exhibits a net compressive strain. LEED is sensitive to these regions, in stark contrast to the He scattering. We have argued therefore for a comparatively reduced LEED sensitivity to the expansions in N domains. The "(½, ½)" form factors for isolated (uncorrelated) N domains are thus anticipated to be much closer to the exact G/2 positions in LEED than in helium scattering.

In other words: scattering He atoms *are* sensitive to a *c(2×2)*-corrugation of the topmost layer of N-containing patches, and are insensitive to subsurface layer positions, or to in-plane compressive relaxations in the selvedge regions of isolated domains. It is the lattice parameter of the individual finite-sized N-containing patches, $a´$, alone – and not the substrate's bulk lattice parameter - that determines the "half-order" He diffraction peak position. In addition a strong correlation of the mean N-N lattice parameter with total N exposures is seen clearly in helium diffraction. Figures 7b and 7c show initially a decreasing lattice parameter with increasing N exposure. The N-containing domains are strictly incommensurate with respect to the substrate bulk.

FIG. 7. (Color online) [100] azimuth room temperature He atom diffraction, from $N^+$ implanted and post annealed Cu(001) surfaces. $E_i = 31.3$ meV. Increasing $N^+$ doses show increasing half-order diffraction intensities at increasing parallel momentum transfers, $|\Delta K_{1/2}|$. $N^+$ doses are color-coded from lowest to highest: black, 50 μC; red, 90 μC; green, 150 μC; blue, 350 μC; cyan, 850 μC; magenta, 1350 μC: and for (c) yellow, 5000 μC. (The highest exposure level curves are omitted in parts (a) and (b) for clarity, to avoid the display of overlapping curves. The zero exposure (brown) point is omitted in part (c) as there is no discernible half-order peak to fit.) **(a)** He diffraction intensity scans, displayed on a logarithmic scale, normalized to the specular diffraction intensity at $\Delta K = 0$. **(b)** Linear-scale normalized-intensity distributions of the $\Delta K_{1/2} = (-½, -½)$ peak region used in evaluation of data for (c). **(c)** 1-D parallel-momentum integrated He intensity of the (-½, ½) peak regions (after background subtraction) *vs.* parallel-momentum positions of intensity peaks, $\Delta K_{1/2}$ (determined from best fit Gaussian curves.) The dashed-dotted line and arrow indicates the movement sense of the data points with increasing N ion exposure.

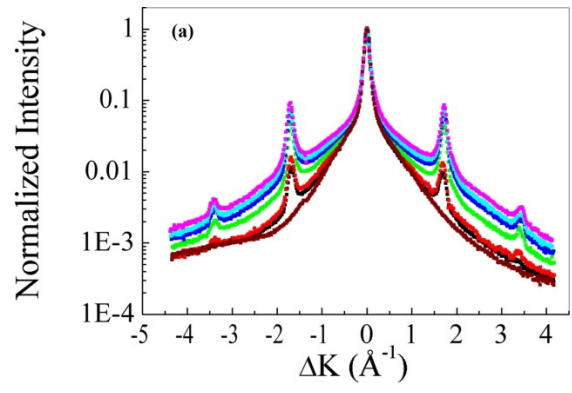

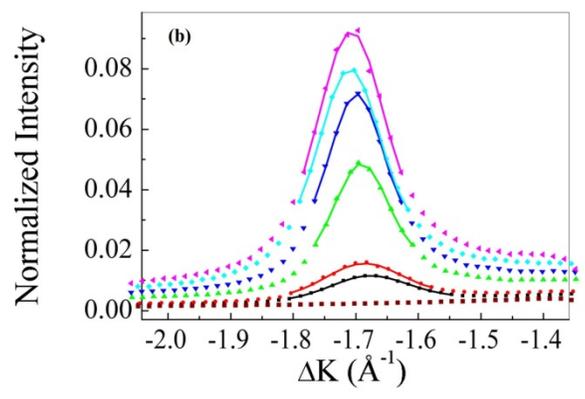

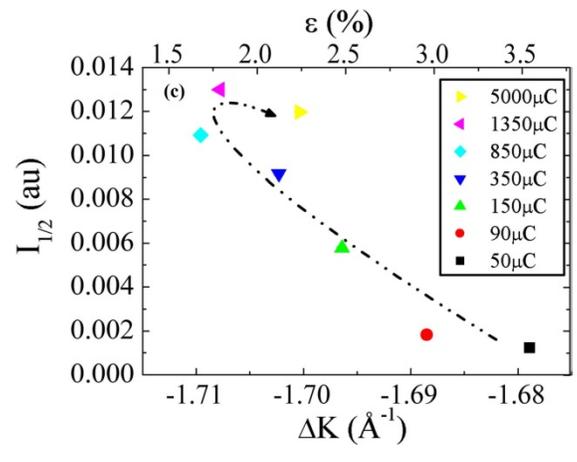

The bulk lattice parameter of 3.61 Å, at room temperature, would dictate a half-order diffraction peak position at $\Delta K$ = 1.74 Å$^{-1}$. The positions of the true measured peaks lie in the range from ~1.68 to 1.71 Å$^{-1}$. At low N coverages an "averaged" N-N nearest-neighbor separation is about 3.74 Å. Stress relief within isolated N domains is manifest in an isotropic surface lattice parameter expansion of roughly 3.5% at the lowest N exposures. This observed strain magnitude is reduced as the domain densities increase at higher N exposures. At around 1000 µC the N-N separations decrease to ~3.67 Å and the expansion virtually halves to 1.8 +/– 0.2%. No evidence is seen for any anisotropic relaxations.

At N$^+$ exposures in excess of 1350 µC, the He diffraction reveals a reversal of the effects of increasing coverage that are evident at lower coverages (i.e., increasing stress levels and decreasing strains). At saturation two effects are seen: the final strain level is increased again to ~ 2.4% and (also shown in Fig 7c) the half-order helium atom diffraction shows a slight reduction in intensity. Both of these observations can be explained by the currently accepted observation on the N/Cu(001) surface, namely that at coverages exceeding 0.35 ML trench-like missing rows aligned along <110> directions are formed.[5] The density of those stress-relieving defects increases until saturation. The missing-row trenches locally will scatter He atoms diffusely and their presence could also modify the corrugation amplitude seen by He in the ordered *c(2×2)* areas. Thus as this defect density increases, the half -order peak intensities may, as is indeed observed, decrease despite the increasing total N levels.

We have found that N$^+$ exposures in excess of 3000 µC produce the saturated phase. We know from STM results that the missing-Cu trenches are first nucleated at lower N$^+$ exposure levels.[5] We suggest then that a flat Cu(001) surface can support the observed 1.8% enlarged N-lattice parameter, but below this critical strain level, i.e. at smaller expansions, the surface stress levels become too large. It appears that the missing-Cu trenches, at increasing N coverages, are initiated at this critical strain level, and presumably at points on the surface where the local compressive stress levels are highest.

Leibsle *et al.* first suggested that the N/Cu may form patches of an incommensurate layer on the Cu substrate.[1,5] A Cu$_3$N crystalline lattice parameter, at 2.69 Å, is 5% larger than that of the substrate. Later LEED studies of a well-ordered surface, at <0.375 ML N, gave precise lateral separations of the N/Cu patches, at 55 Å.[26] On the same surface, given a lack of asymmetry in satellite spots around an *(hk)* bulk diffraction peak, Sotto et al concluded that Cu and N overlayer strains were below 0.05%.[26] Yet channeling and blocking measurements, in combination with an atomistic simulation of inhomogeneous stresses, with an assembly of stressed N-on-Cu patches indicated that Cu lateral displacements were as much as ~ 0.35 Å, although the rms displacement of surface Cu atoms was as low as 0.15 or 0.16Å.[15] The strain on this surface, averaged across a patch, was about 2.3%. The quenched molecular dynamics simulations showed also that the surface atomic species at the edge of a patch were subject to larger displacements than those at the center of a 2-D patch. A very similar strain displacement pattern also gave excellent agreement with X-ray diffraction rod intensities.[16] The interpretation of the X-ray diffraction, however, is not simple and direct, as the diffraction pattern is influenced by the deep ($\gtrsim 50$Å) stress field experienced below each patch.

Careful analysis of precise STM measurements of N-atom displacements showed directly that the strain patterns in nearly isolated N patches are inhomogeneous.[10,11,27] It was concluded that an rms displacement within an N-containing patch is as large as 0.6Å and that the maximum displacements typically do not exceed one half of a Cu-Cu nearest-neighbor spacing, 2.55 Å. It was also concluded that N does, on average, sit in fourfold hollow sites and that the largest displacements are seen at the boundaries, i.e., at the edges of patches.

The quantitative analysis of distortions in N on Cu arrays with STM has proven difficult because of possible electronic effects and anisotropy in the scanning tip. Arguably diffraction gives rise to data that are more representative of many patches. The measured X-ray reciprocal lattice rod intensities fit well with those calculated for a simulated non-uniform strain pattern at higher N surface concentrations  Although the He diffraction presented here does not yield information about the non-uniformity of strain and stress levels within N-containing patches, it does reveal variations of strain levels between surfaces within a wide range of $N^+$ exposures.

All data to date are consistent with the view that a *c(2×2)* N/Cu(001) phase region is under compressive stress.  The stress is in part relieved through expansion of the mean near surface-Cu lattice parameter, and possibly through N displacements with respect to the surface-Cu defined hollow sites.  We have now shown as well that the degree of stress relief is dependent on the N coverage.  Surface strains can be reduced to < 2%.  In contrast, at the lowest coverage we have seen a lateral expansion as high as 3.5%.  The observed range of average strain magnitudes is fully consistent with Croset *et al.*'s calculations.[16]

### B. Results from DFT Calculations

#### 1. Structural relaxations of the *c(2×2)* N/Cu(001) phase

In Table 1 we present the results of our investigation into structural relaxation within an *ideal c(2×2)*-N overlayer on the unreconstructed Cu(001) system and make comparisons with available theoretical and experimental results. Recall the top view of the *c(2×2)*-N phase as shown in Fig. 1a. The vast majority of experimental and theoretical studies (see Table 1), agree that N atoms adsorb in fourfold hollow sites on Cu(001) without any distortion of the substrate. Although the N adsorption height varies from 1.5 Å to zero above the upper Cu-atom plane, most studies put it between 0.1 Å and 0.6 Å. Our calculations for the *ideal c(2×2)*-N/Cu(001) surface also predict that N adsorbs 0.17 Å above the first-layer Cu atoms, i.e., N atoms are almost coplanar with the outermost layer of Cu atoms. They also agree on a large expansion of the first Cu interlayer spacing with respect to the bulk: $\Delta_{12}$ varies from 15% to 5% depending on the techniques used, recent DFT-GGA studies find it to lie in the range of 7.7 and 9.1%. The relevant experimental structural parameters for the proposed clock and rumpling models, discussed in Sec. I, are presented in Table 2. The total displacement of a Cu atom, in the clock model (illustrated in Fig. 1c) is reported to be 0.14 Å.[6] For the rumpling model the difference in height between the upper and lower Cu atoms  in the top layer was reported to be 0.34 Å.[7,8] As a result, N atoms were to locate between the upper and lower Cu atoms, at 0.07 Å below the upper half-plane of Cu atoms.

Table. 1. Calculated structural parameters of ideal c(2×2)-N/Cu(001) compared with available theoretical and experimental results for an undistorted substrate. For the definition of the structural parameters used in this table, refer to Fig. 6.

| Method | $d_{N-Cu1}$ (Å) | $r_1$ (Å) | $\Delta_{12}$ (%) | $\Delta_{23}$ (%) | Reference |
|---|---|---|---|---|---|
| LEED | 1.45 | 0 | | | 28 |
| LEED | 1.46 | 0 | | | 29 |
| LEED | 0.6 | 0 | | | 30 |
| LEED | 0.0 | 0 | +7.7 | | 6 |
| SEXAFS | 0.41 | 0 | | | 31 |
| SEXAFS | 0.4 | 0 | +4.7 | +0.3 | 32 |
| HF Cluster Model | 0.36 | 0 | | | 33 |
| HF Cluster Model | 0.6 | 0 | | | 34 |
| DFT-GGA | 0.48 | 0 | | | 35 |
| Helium Ion Channelling | | | +15.0 | 3.0 | 15 |
| X-ray Diffraction & Molecular Dynamics | | | +14.0 | 1.5 | 16 |
| DFT-GGA | 0.21 | 0 | +7.7 | +0.5 | 13 |
| grazing incidence X-ray diffraction | 0.15 | 0 | +14.0 | +1/5 | 36 |
| DFT-GGA | 0.2 | 0 | +9.1 | +0.9 | 37 |
| DFT-GGA | 0.17 | 0 | +7.8 | +0.2 | This study |

Table. 2. Structural parameters of hypothetical clock- and rumpling- reconstructed N/Cu(001) surfaces from experiment and theory. DFT-X1 and DFT-X2 (X=C,R) represent, respectively, theoretical structures for clock (X=C) and rumpling (X=R) models that were investigated in this study. They were obtained by fixing N or Cu atoms in the top layer to positions either close to experimental ones (DFT-X1) or ones based on the analogy with that of other system (DFT-X2) while relaxing all other atoms.

|  | Clock model | | | Rumpling model | | |
|---|---|---|---|---|---|---|
| Method | LEED[a] | DFT-C1 | DFT-C2 | PhD[b]/STM[c] | DFT-R1 | DFT-R2 |
| $d_{N-Cu_1}$ (Å) | 0.06 | 0.21 | 0.05 | -0.07 | -0.07 (fixed) | 0.10 (fixed) |
| $d_{N-Cu_2}$ (Å) | 1.91 | 2.13 | 2.14 | 1.99 | 2.09 | 2.13 |
| $r_1$ (Å) | - | - | - | 0.34 | 0.34 (fixed) | 0.24 (fixed) |
| $\delta$ (Å) | 0.14 | 0.14 (fixed) | 0.42 (fixed) | - | - | - |
| $\Delta_{12}$ (%) | +2.5 | +9.6 | +13.7 | +4.7 | +8.8 | +3.8 |
| $\Delta_{23}$ (%) | -2.5 | +0.5 | +0.7 | +0.3 | -1.4 | -3.2 |
| $r_2$ (Å) |  | 0.12 | 0.05 | - | - | - |

[a] Ref. 6
[b] Ref. 7
[c] Ref. 8

To check on the stability of the proposed clock, denoted by C, or rumpling, denoted by R, model structures, we first generated the model structures as deduced by other workers and compare these with second alternative structures, as described below. Table 2 reports on our results for models DFT-X1 and DFT-X2, where X = C or R depending on the model in question. The numbers 1 and 2 refer to the two models of each type.

For the clock model, we first carry out DFT-C1 calculations to examine its energetics using the lateral-shift parameter ($\delta$ in Fig. 6) as described by Zeng, *et al.*[6] Specifically, we fix the lateral displacement of Cu atoms in the first layer at 0.14 Å and allow all other atoms to relax completely. To pinpoint the structural and energetic effects of this lateral-shift magnitude, we then increase $\delta$ to 0.42 Å and again allow all other atoms to relax. (See DFT-C2 in Table 2.) The most notable structural changes between the relaxations of DFT-C1 and DFT-C2 occur in the N height ($dN-Cu_1$), which reduces from 0.21 to 0.05 Å, and the first-layer Cu expansion ($\Delta_{12}$), which increases from 9.6 to 13.7%. As Cu atoms shift more, N atoms can more closely approach the surface, implying that the top layer interacts most strongly with the N atom. However, we find that both structures are *unstable* since they always return to the undistorted *c(2×2)*-N/Cu(001) structure during relaxation when the initial constraint (δ for each model) is lifted. Since DFT-C2 has a higher total energy than DFT-C1, the degree of instability increases as the shift magnitude increases.

Analogously, for the rumpling model, we begin with DFT-R1 calculations of a rumpled surface with fixed rumpling displacements; We fix the amplitude of the rumpling level ($r_1$ in Fig. 6) of the Cu atoms in the first layer at 0.34 Å and the N atoms at 0.07 Å below the atoms in that layer.[7,8] (See DFT-R1 in Table 2.) Relaxation of this experimentally suggested structure gives a first-layer expansion $\Delta_{12}$ of 8.8%, which is comparable to that of the unrumpled (ideal) structure. But this structure, too, is unstable, as its total energy is higher than that of the unrumpled one: N atoms are always pushed upwards and the rumpling of Cu atoms systematically disappears, once the initial constraints are lifted.

To illustrate the effects from the two parameters derived from experiment, we carry out DFT-R2, with an increased N height, at 0.1 Å above the upper layer Cu atoms and a decreased rumpling at 0.24 Å. We can think of DFT-R2 as calculations from a state which is intermediate between an ideal, unrumpled surface and that assumed in DFT-R1. We find that the effect of this relative positioning of N atoms is to notably decrease $\Delta_{12}$ from 8.8 to 3.8%. The fact that DFT-R2 has a higher energy than DFT-R1 might suggest that DFT-R1 could be a local minimum. It is not, however, because in absence of the constraints described above, it always returns to the unrumpled ideal *c(2×2)*-N structure. All these facts suggest that even in terms of stability neither model – clock or rumpling – is favorable for *c(2×2)*-N/Cu(001) surfaces.

It might be wondered, however, whether the surface might undergo rumpling in a more realistic situation, specifically, at sub-saturation coverage, where the N-overlayer density concentrates in localized patches. Hence, it is necessary to check whether a stripe can induce rumpling as large as was proposed in the rumpling model. We present the calculated structural parameters of these striped surfaces for the in-phase boundaries in Table 3 and for the out-of-phase boundaries in Table 4. Yoshimoto et al.[14] have already investigated the relaxations of striped $c(2\times 2)$-N/Cu(001) surfaces of N patch width $l = 5, 6, 8\ a_o$ and stripe widths $d = 1, 3\ a_o$ – larger than those under examination in our study.

Table 3. Theoretical structural parameters of the clean and striped surfaces of N/Cu(001) with in-phase boundaries with different N-patch width $l$ and stripe width $d$.

| $l\ (a_o)$ | $d\ (a_o)$ | unit cell | $\overline{dN\text{-}Cu_1}$ (Å) | $\overline{r_1}$ (Å) | $\overline{\delta}$ (Å) | $\overline{\Delta_{12}}$ (%) | $\overline{dCu_1\text{-}Cu_1}$ (%) | $\overline{dN\text{-}N}$ (%) |
|---|---|---|---|---|---|---|---|---|
| 0 | ∞ | $p(1\times 1)$ | N/A | 0 | 0 | -5.4 | 0 | 0 |
| 1 | 1 | $(2\sqrt{2}\times\sqrt{2})R45°$ | 0.129 | 0.074 | N/A | +2.0 | +2.5 | N/A |
| 3 | 1 | $(4\sqrt{2}\times\sqrt{2})R45°$ | 0.144 | 0.125 | 0.087 | +5.7 | +2.2 | 3.2 |
| 3 | 2 | $(5\sqrt{2}\times\sqrt{2})R45°$ | 0.140 | 0.165 | 0.074 | +3.9 | +2.2 | 3.1 |
| 4 | 1 | $(5\sqrt{2}\times\sqrt{2})R45°$ | 0.143 | 0.106 | 0.084 | +6.5 | +1.9 | 2.7 |

Table 4. Theoretical structural parameters of $c(2\times 2)$ N/Cu(001) surfaces with out-of-phase boundaries for different N-patch width $l$ and stripe width $d$.

| $l\ (a_o)$ | $d\ (a_o)$ | unit cell | $\overline{dN\text{-}Cu_1}$ (Å) | $\overline{r_1}$ (Å) | $\overline{\delta}$ (Å) | $\overline{\Delta_{12}}$ (%) | $\overline{dCu_1\text{-}Cu_1}$ (%) | $\overline{dN\text{-}N}$ (%) |
|---|---|---|---|---|---|---|---|---|
| 3 | 1.5 | $\begin{pmatrix} 5 & 4 \\ 0 & 1 \end{pmatrix}$ | 0.153 | 0.168 | 0.091 | +4.5 | +2.2 | +3.0 |
| 4 | 0.5 | $\begin{pmatrix} 5 & 4 \\ 0 & 1 \end{pmatrix}$ | 0.181 | 0.113 | 0.064 | +6.3 | +1.3 | +1.5 |

An N patch (width $l=4\ a_o$) in stripe formation induces a substantial expansion of the first-to-second interlayer distance $\Delta_{12}$ of 6.5%, not greatly at variance from $\Delta_{12}$ in the ideal $c(2\times 2)$ N overlayer (7.8%). Thus, if the N-patch width is sufficiently larger than the stripe width, the N-patch functions more like the ideal (infinite) N overlayer, so that the effect of a stripe or a trench becomes weak. This is also clearly seen in the average lateral $Cu_1$-$Cu_1$ distances, $\overline{dCu_1\text{-}Cu_1}$, which decreases as $l$ increases. Conversely, if the patch width is relatively smaller, the effect of a stripe is stronger, as can be seen in the relaxations of the top layer: $\Delta_{12}$ is just 2.0% for a stripe of $l = 1\ a_o$ and $d = 1\ a_o$. (See Table 3).

In connection with the rumpling model, the displacements of interest are the vertical ones of Cu atoms in the top layer, which we present in Fig. 8 for the surface structure ($l = 4\ a_o$, $d = 1\ a_o$). They differ according to location with respect to stripe. Inner Cu atoms within the patch in the first Cu layer show a small rumpling ($r_1$ in range of 0.04 − 0.1Å), which decreases even further as the patch width $l$ increases. The most significant rumpling appears among Cu atoms at the patch edge adjacent to stripe, as shown in Fig. 8. Thus, the effect of any given stripe is limited to the relaxation of the Cu atoms at the edge, and quickly dies away towards inner atoms. Therefore, we expect that for a much larger island, such as the one observed in experiment (55 Å × 55 Å), any rumpling of inner Cu atoms must be negligible. Besides, our calculations show that N atoms always sit above Cu atoms by at least 0.1 Å for larger values of $l$ − in contrast to the proposed rumpling model, within which the N atoms are either above what it characterizes as a lower Cu-atom half plane or below those it takes to be above. We will discuss these results further below, in our treatment of surface stress.

FIG. 8. Calculated vertical displacements of Cu atoms in the top layer for the surface structure ($l = 4\ a_o$, $d = 1\ a_o$) shown in Fig. 2d. The displacements here are specified with respect to the average height of the first Cu layer.

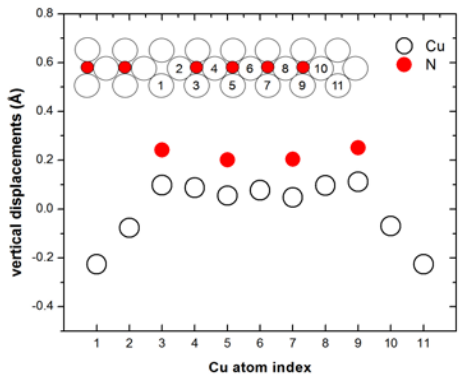

## 2. Surface stresses of clean Cu(001), ideal N/Cu(001), and clock- and rumpling-reconstructed N/Cu(001) surfaces

Table 5. Calculated surface stress for clean, ideal c(2×2)-N, clock-reconstructed, and rumpled Cu(001) surfaces.

| Surface | Unit cell | Analytical stress (N/m) | Numerical stress (N/m) | Other study (N/m) |
|---|---|---|---|---|
| clean | p(1×1) | +1.27 | +1.31 | +1.4[a], +1.51[b],+1.38[c] +2.10[d] |
| ideal c(2×2)-N | c(2×2) | -4.20 | -4.21 | -5.3[a] -4.0 ~ -4.2[e] |
| rumpling (DFT-R1) | c(2×2) | -5.24 | -5.47 | |
| rumpling (DFT-R2) | c(2×2) | -6.55 | -6.68 | |
| clock (DFT-C1) | p(2×2) | -3.36 | -3.61 | |
| clock (DFT-C2) | p(2×2) | -1.18 | | |

[a] Ref. 14. DFT-GGA  [b] Ref. 39. DFT-GGA  [c] Ref. 40. EAM  [d] Ref. 41. Modified EAM.
[e] Ref. 36. Tight-binding approximation. Surface stress value for clean Cu(001) is assumed to be in range of 1.3 and 1.5 N/m.

We have calculated surface stresses for several surface configurations introduced earlier. These include the hypothetical surface structures of the rumpled and clock-reconstructed N/Cu(001) surfaces. We present the results in Table 5, in which a negative value means compressive stress (a tendency to expand the surface area), while a positive value implies tensile stress (a tendency to contract the surface area). Analytical and numerical surface stresses are in excellent agreement, within a maximum deviation of 0.28 N/m. More importantly, our results square well with those of other studies, both theoretical and experimental. (We performed numerical stress calculations selectively as a check for our analytical stress results. From now on we report the analytical stress results if not otherwise specified.) Our calculated surface stresses for clean Cu(001) and *c(2×2)*-N on unreconstructed Cu(001) surfaces are 1.3 and -4.2 N/m, respectively. Yoshimoto and Tsuneyuki reported 1.4 and -5.3 N/m for these surfaces, respectively.[14] The figures for clean surface are nearly identical, but our value for *c(2×2)*-N/Cu(001) is 26% smaller than their analogous result. (See Table 5.) On the other hand, the surface stress difference between the N patch and stripe regions reported in previous studies, are 5.5 N/m,[36] 6.1 N/m[15] and 7.0 N/m.[16] While our calculated stress change (5.5 N/m) is in excellent agreement with that of Prevot et al. (5.5 N/m)[36], all these studies[14-16,36] unanimously agree that N overlayer induces a large stress change, which happens to be in the range of -5.5 and -7.0 N/m.

Now we discuss the surface stress levels of clock and rumpling models for the *c(2×2)*-N/Cu(100) structure. Applying the constraints discussed in Sec. III.B.1, we calculate the surface stress for the two rumpling model structures DFT-R1 and DFT-R2 to be -5.24 and -6.55 N/m, respectively (See Table 5). These values constitute a jump of 1.04 − 2.35 N/m beyond those of the unrumpled ideal surface. That is, the rumpling model turns out not to relieve stress, but indeed to intensify it.

For the clock model, the calculated surface stresses for the two structures DFT-C1 and DFT-C2 are -3.36 and -1.18 N/m, respectively. Recall from Sec. III.B.1 that for DFT-C1 we chose a clock shift value (0.24 Å) between the two experimental values reported for lateral displacements (0.14 Å[6] and 0.28 Å[7]), while for DFT-C2 we chose a value (0.42 Å) comparable to that for C/Ni(001) (0.4 Å). The stress reduction caused by the rotation from undistorted (ideal) Cu(001) increases as the rotation increases. It is striking to find the large surface stress on the undistorted surface (-4.2 N/m) is substantially relieved by a shift of 0.42 Å in DFT-C2. This demonstrates that the clock displacements do indeed contribute to relief of the compressive stress. Nevertheless, recall that this clock displacement is not energetically favorable. Just as for the rumpling models, as the displacement increases the total energy increases. The total energy increase in DFT-C2 is related to the reduced rumpling in the *second* Cu layer ($r_2$ in Table 2) from 0.12 Å (DFT-C1) to 0.05 Å (DFT-C2), which registers the interaction strength between N and the Cu atom directly below it. Both the N-$Cu_2$ bond and the N-$Cu_1$ bond are weakened with Cu rotation (i.e., the bond lengths increase,) resulting in an increase of the total energy. These results remind us of the importance of considering both stress-relief and energetic arguments in discussion of stress-related phase transitions on this and other surfaces. Although various stress-relief mechanisms can be imagined, only the energetically-favorable ones will occur.[38,42] For example, though the clock displacement would substantially relieve surface stress for C/Ni(001), O/Ni(001) and N/Cu(001),[38] it actually takes place only in C/Ni(001), and not in O/Ni(001) nor in N/Cu(001).[43]

### 3. Surface stress of the *c(2×2)* N/Cu(001) surfaces with stripes

The stress-relief mechanisms active in N/Cu(001) system are generally considered to involve variations in surface density resulting from different spatial periodicities. The stress is in part relieved through expansion of the mean lattice parameter of N-containing regions. We find that the degree of stress relief is dependent on N coverage: at high N coverages surface strains are <2%, whereas at low coverage we have seen an N-N lateral expansion as high as 3.5%. Thus the following competing stress-relief mechanisms may be at work on the N/Cu(001) surface:

(a) At low N coverage the clean surface stripes which coexist with *c(2×2)* patch-like structures may be the leading cause of stress relief.[1]

(b) At somewhat higher coverage stress relief mechanism may involve the formation of missing copper rows in the clean surface regions.[3]

(c) At saturation coverage (disappearance of clean surface region) stress relief may come through the formation of Cu-vacancy trenches along the <110> direction beneath the N overlayer.[11]

We present the results of our calculations, in the light of the above. As indicated in Figs. 2 and 3, on striped surfaces N patches and stripes are alternating. The dimension of the corresponding surface unit cells, *l+d*, represents the spatial periodicity of stripes at the surfaces. Thus, smaller *l+d* (or simply *l* if *d* is fixed) corresponds to larger stripe density at the surface. (*l = 0* would mean zero-width N-patch – i.e., clean surface.)

FIG. 9. Calculated stress levels for striped surface phases with respect to N-patch width *l* for in-phase stripe boundaries. The surfaces considered in this graph have stripe widths, *d*, of *1 $a_o$* only.

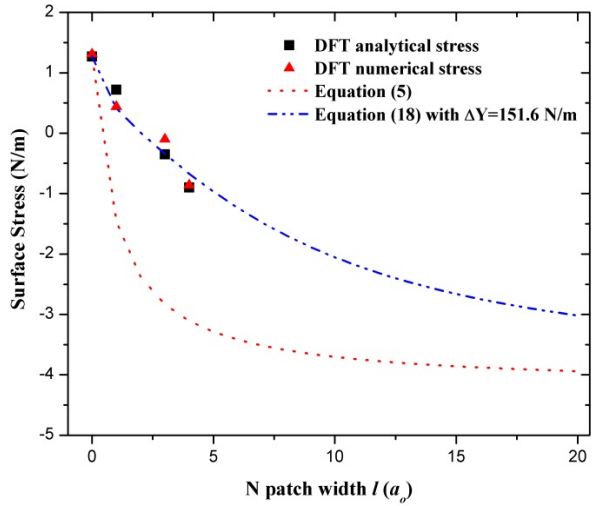

Table 6. Calculated surface stress for striped surfaces with different N-patch width *l* and stripe width *d*.

| Boundary type | Unit cell | $l$ ($a_o$) | $d$ ($a_o$) | Analytical stress (N/m) | Numerical stress (N/m) |
|---|---|---|---|---|---|
| in-phase | $(2\sqrt{2}\times\sqrt{2})R45°$ | 1 | 1 | +0.72 | +0.44 |
|  | $(4\sqrt{2}\times\sqrt{2})R45°$ | 3 | 1 | -0.35 | -0.1 |
|  | $(5\sqrt{2}\times\sqrt{2})R45°$ | 4 | 1 | -0.90 | -0.86 |
|  | $(5\sqrt{2}\times\sqrt{2})R45°$ | 3 | 2 | -0.30 |  |
| out-of-phase | $\begin{pmatrix}5 & 4\\0 & 1\end{pmatrix}$ | 3 | 1.5 | -0.38 | -0.21 |
|  | $\begin{pmatrix}5 & 4\\0 & 1\end{pmatrix}$ | 4 | 0.5 | -2.24 | -2.51 |

In striped surfaces both compressive and possibly even tensile stress regions can coexist since stripes exhibit different stress levels from those of the patches. As we reduce *l* gradually to zero keeping $d = 1\ a_o$ (i.e., we systematically add more stripes with the uniform stripe width *d*), we would expect that the compressive stress in the substrate reduces, eventually, to the level within the tensile clean Cu(001) surface at $l = 0$. In Table 6 we report our results of stress levels averaged over the unit cell, for striped surfaces with N patch widths $l = 4\ a_o$ to $l = 0$ and stripe width $d = 1\ a_o$ and we show selected results in Fig. 9 for the in-phase boundary. Clearly, striped surfaces exhibit remarkable stress relief, and as N-patch width *l* *decreases* the stress relief *increases*. For small *l* only, our analytical calculations of surface stress suggest that the stress relief is approximately

27 | P a g e

proportional to $1/l$.

To extrapolate our results for large $l$, we develop a model based on the fact that, for a regularly striped surface with N-patches of width $l$ and stripes of width $d$, surface stress receives contributions from both. Therefore, for a striped surface we have

$$\sigma_{avg}(l,d) = \sigma_{patch}\frac{l}{L} + \sigma_{stripe}\frac{d}{L}, \tag{4}$$

where $\sigma_{patch}$ and $\sigma_{stripe}$ are surface stresses, averaged respectively throughout the N-patch and the stripe region and $L$ is the surface length (1D unit cell size) equal to $l + d$. According to equation (4) $\sigma_{avg}$ approaches $\sigma_{patch}$ as $l \to \infty$ but approaches $\sigma_{stripe}$ as $l \to 0$. As an initial guess, *if* we assume that $\sigma_{patch}$ and $\sigma_{stripe}$ do not change from their initial values ($\sigma_{patch}^{initial}$ = -4.2 N/m and $\sigma_{stripe}^{initial}$ = +1.27 N/m) regardless of $l$, the resultant surface stress, $\sigma_{avg}$, would be

$$\sigma_{avg}^{initial}(l,d) = \sigma_{patch}^{initial}\frac{l}{L} + \sigma_{stripe}^{initial}\frac{d}{L}. \tag{5}$$

This hypothetical stress is presented by the dotted curve in Fig. 9. This stress turns out to be much larger in magnitude (more compressive) than the calculated ones (See Fig. 9). This large discrepancy arises because Equation (5) does not include the stress relief in $\sigma_{patch}$ and $\sigma_{stripe}$ contributed by the *relaxations of first-layer $Cu_1$-$Cu_1$ bond lengths* at the surface as a result of *the formation of stripes*. In reality, $\sigma_{patch}$ and $\sigma_{stripe}$ should change from their initial values. As a first order approximation, these changes can be expressed as follows:

$$\sigma_{patch}(\lambda^{patch}) = \sigma_{patch}^{initial}(\lambda_0^{patch}) + \left(\frac{d\sigma_{patch}}{d\lambda^{patch}}\right)_{\lambda^{patch}=\lambda_0^{patch}} (\lambda^{patch} - \lambda_0^{patch}) \tag{6}$$

and

$$\sigma_{stripe}(\lambda^{stripe}) = \sigma_{stripe}^{initial}(\lambda_0^{stripe}) + \left(\frac{d\sigma_{stripe}}{d\lambda^{stripe}}\right)_{\lambda^{stripe}=\lambda_0^{stripe}} (\lambda^{stripe} - \lambda_0^{stripe}), \tag{7}$$

where $\lambda_0^{patch}$ and $\lambda^{patch}$ are the initial and final $Cu_1$-$Cu_1$ bond lengths of the first-layer Cu substrate averaged beneath the N patch, while $\lambda_0^{stripe}$ and $\lambda^{stripe}$ are the corresponding variables averaged within the Cu stripe. As a result, within this 1-D model with uniform but distinct patches and stripes, the stress changes $\Delta\sigma = \sigma - \sigma^{initial}$ will be:

$$\Delta\sigma_{patch} = Y^{patch} \varepsilon^{patch} \qquad (8)$$

and

$$\Delta\sigma_{stripe} = Y^{stripe} \varepsilon^{stripe} \qquad (9)$$

, respectively, where $\varepsilon = \frac{(\lambda - \lambda_0)}{\lambda_0}$ is the strain – i.e., expressible as the fractional expansion of the (average) first-layer Cu-Cu bond length $d(Cu_1\text{-}Cu_1)$, with respect to the substrate lattice parameter $\lambda_o = a_o$ – and

$$Y^{patch} = \left(\frac{d\sigma_{patch}}{d\varepsilon^{patch}}\right)_{\lambda_0} \qquad (10)$$

$$Y^{stripe} = \left(\frac{d\sigma_{stripe}}{d\varepsilon^{stripe}}\right)_{\lambda_0} \qquad (11)$$

are (microscopic) *area-averaged* elastic moduli of the first-layer Cu substrate beneath the N patch and at the Cu stripe, respectively. Here we assume that these moduli are fixed, and depend solely on the presence or absence of the adsorbed nitrogen. The expansion strain $\varepsilon$, under a very simplifying assumption, is taken to be uniform within a patch and must be one function of $l$ and $d$. Also, since $(L\lambda_o)/a_o = (l\lambda^{patch} + d\lambda^{stripe})/a_o$ is conserved,

$$\varepsilon^{stripe} d = -\varepsilon^{patch} l \qquad (12)$$

As a result, the final form of Equation (4) will be,

$$\sigma_{avg}(l,d) = \sigma_{avg}^{initial}(\lambda_0) - \Delta Y \frac{d}{L} \varepsilon^{stripe}, \qquad (13)$$

where $\Delta Y = Y^{patch} - Y^{stripe}$. Note that stress relief $\Delta\sigma = \sigma_{avg} - \sigma_{avg}^{initial}$ is linearly proportional to strain $\varepsilon$ for a given $l$ and $d$. Note also that while $\sigma_{avg}^{initial}$ in Equation (5) is in fact just a function of one variable $l/d$ alone, we are not certain of the dependence on $l/d$ of $\sigma_{avg}(l,d)$ in Equation (13), since the exact dependence of $\varepsilon^{patch}$ and $\varepsilon^{stripe}$ upon $l$ and $d$ is unfortunately not known. Nevertheless we attempt here to estimate these functional forms given some self-evident facts and further assumptions as follows. First, as $l \to \infty$, we require that $\varepsilon^{patch} \to 0$ since there should be no lateral expansion in an extended 0.5 ML N/Cu(001) patch. Second, each will vary monotonically as the N coverage is varied. In consideration of the $d = 1$ stripe calculations reported upon in Tables 3, and 6, and in Fig. 9, we thus propose that for fitting purposes $\varepsilon^{stripe}$ and $\varepsilon^{patch}$ can take the following functional forms:

$$\varepsilon^{stripe}(l) = -a\,(1-\exp(-bl^c)) \tag{14}$$

$$\varepsilon^{patch}(l) = -\frac{\varepsilon^{stripe}(l)}{l} = \frac{a\,(1-\exp(-bl^c))}{l} \tag{15}$$

The parameters $a$, $b$, and $c$ can be found by fitting to our calculated values of surface averaged lateral expansions within patches. We have utilized the signs such that $a$, $b$ and $c$ are expected to take positive values. For the special case with $l = 1$, both strains are necessarily equal and opposite, and independent of the variable $c$;

$$\varepsilon_{patch} = -\varepsilon_{stripe} = 0.025 = a(1-\exp(-b)) \tag{16}$$

The above numerical value is taken from Table 3. Other numerical ($d = 1$) strain data from this table is then fitted using,

$$\varepsilon^{patch}(l) = -\frac{0.025\,(1-\exp(-bl^c))}{(1-\exp(-b))l} \tag{17}$$

From $\varepsilon^{patch}$, which is 2.5, 2.2 and 1.9% for $l = 1, 3, 4\ a_o$ respectively, (c.f. $\overline{dCu_1 - Cu}_1(\%)$ in Table 3,) we obtained the best fit parameters $a = 0.128$, $b = 0.214$, and $c = 1.1$. Note that Equation (15) guarantees that as $l \to \infty$, $\varepsilon^{stripe} \to -a$, that as $l \to 0$, $\varepsilon^{stripe} \to 0$, and that $\varepsilon^{stripe}$ varies monotonically with $l$ as required. Given that this fit is for $d = 1$ stripe widths, then $a$ also represents the maximum achievable patch extension, which is $0.128\ a_o$. Equation (17) guarantees that as $l \to \infty$, $\varepsilon^{patch} \to 0$, and that as $l \to 0$, $\varepsilon^{patch} \to 0$. But with the choice of $c > 1$, the patch strain can be seen to increase unphysically with increasing $l$ at small $l$. This unphysical solution at $l < 1$ is an artifact of our continuum model, in which we treat $l$ as a continuous variable while in reality $l$ could take only the multiples of lattice constant $a_o$.

Substituting the model form of (14), and the definition of (5), in the 1-D model result (13) we obtain

$$\sigma_{avg} = \frac{l\,\sigma^{initial}_{patch} + \sigma^{initial}_{stripe} + \Delta Y\ a\,(1-\exp(-bl^c))}{l+1} \tag{18}$$

Clearly if $l$ is 0, $\sigma_{avg} = \sigma_{stripe}^{initial} = 1.27$ N/m, and as $l \to \infty$, $\sigma_{avg} = \sigma_{patch}^{initial} = -4.2$ N/m. In addition, by fitting the surface stresses for striped surfaces ($\sigma = 1.27, 0.72, -0.35, -0.9$ N/m for $l = 0, 1, 3, 4\ a_o$, respectively, in Table 6) and by using Equation (17), we find that $\Delta Y = 151.6$ N/m. The values derived here, of course, cannot be expected to precisely represent true values of a real surface since we have made no attempt to model the 2-D array structure of patches or to consider spatial variations of stresses and strains. On the other hand, the 1-D moduli, $Y^{patch}$ and $Y^{stripe}$, represent the stiffness of each of the surfaces with respect to that of the substrate material. Either one or both of these moduli may therefore even be negative in value. Suffice it to say that our determined difference value, $\Delta Y = 151.6$ N/m, implies that the surface of the patch has a stiffer modulus than that of the striped region surface. This observation might be anticipated from consideration of the higher near-surface packing density with the included/added N atoms. The fact that N-containing patch dimensions (~55Å) are virtually independent of surface coverage would also be supportive of a strong asymmetry in the relative stiffness of the surface regions. Finally, the results of Ng and Vanderbilt[43] imply that for assembly of a surface without a modulus asymmetry the spatial distribution of N patches at low N coverages, $\theta_N$, should be equivalent to the spatial distribution of the other (N-free) phase at a complementary N coverage $\theta_{free} = 0.5 - \theta_N$. No such equivalence is observed: e.g. N-containing patches tend to be square, while only much smaller, diamond (rotated,) N-free patches tend to form at the intersections of N-free stripes. The observed inequivalence of N-containing and N-free spatial distributions is thus also supportive of our suggested asymmetries in surface region moduli.

Finally we can obtain the model $\sigma_{avg}(l)$ which we display by the dash-dot curve in Fig. 9. Apparently, from our model form of $\sigma_{avg}$, as the stripe density increases ($l$ decreases), compressive stresses within the N-patches reduce monotonically. This stress relief is achieved by means of expansions within the N-patches and contractions within the stripe. Thus, the formation of the stripe boundary is critical in stress relief. On the other hand, as the stripe density decreases ($l/d$ increases), stress reduction is not so effective. At the experimental length of N-patches ($l \approx 15$), $\sigma_{avg}$ is 2.75 N/m, *just 65% of the level of the ideal c(2x2) N/Cu(001) surface.* The slow stress reduction at large $l/d$ is because lateral relaxation $\overline{dCu_1 - Cu_1}(\%)$ is subjected to Equation (12). Accordingly, at large N-patch width, stripe width is critically important in stress reduction.

We now discuss the striped surfaces with out-of-phase boundaries. As observed in experiment,[2,3] the N-patch alignment in the boundary is not unique: it could be either in phase or out of phase. Figure 3 shows the out-of-phase boundaries with different stripe widths and Table 6 presents the calculated stress. Comparison of the stress levels for the out-of-phase boundaries with those for the in-phase boundaries is illuminating. We see a large stress relief from the stripe width of 0.5 $a_o$ for $l=4$ $a_o$ to that of 1.5 $a_o$ for $l=3$ $a_o$ the latter showing ~ 2 N/m more reduction than the former. This remarkable effect of stripe width in stress reduction is not obtained for in-phase boundaries, where as we increase $d$ to 1 $a_o$ for $l=4$ $a_o$ from 2 $a_o$ for $l=3$ $a_o$ the increased stripe width gives a fractional reduction of only 0.6 N/m. On the other hand, even taking into account the smaller stripe thickness in the out-of-phase boundary, the stress level for the out-of-phase boundary with stripe width of 0.5 $a_o$ is certainly much more compressive than that for the in-phase boundary with stripe width of 1.0 $a_o$ for $l=4$ $a_o$: -2.24 vs. -0.9 N/m, respectively. However, this is not true for the larger stripe widths, namely, of 1.5 $a_o$ (out-of-phase) and 2.0 $a_o$ (in-phase) for $l=3$ $a_o$: -0.38 vs. -0.30 N/m, respectively: instead, these exhibit similar stress levels. Thus our calculations show relative effectiveness of the in-phase boundary for the *monoatomic* stripe, confirming a model recently proposed by Yamada *et al.* based on STM measurements.[3] Overall, stress relief in striped surfaces strongly depends on the local geometry of the boundary.

### 4. Surface stresses of the *c(2×2)* N/Cu(001) surfaces with a missing-Cu row boundary

Table 7. Calculated surface stress for surfaces with a missing-row boundary.

| Surface phase | Unit cell | $l$ ($a_o$) | $d$ ($a_o$) | Analytical stress (N/m) | Numerical stress (N/m) |
|---|---|---|---|---|---|
| monoatomic-wide boundary: along ⟨100⟩ | $(5\sqrt{2}\times\sqrt{2})R45°$ | 4 | 1 | +0.83 | +1.26 |

Table 8. Theoretical structural parameters of c(2×2) N/Cu(001) surfaces with a missing-row boundary

| $l\ (a_o)$ | $d\ (a_o)$ | unit cell | $\overline{dN\text{-}Cu_1}$ (Å) | $\overline{r_1}$ (Å) | $\overline{\delta}$ (Å) | $\overline{\Delta_{12}}$ (%) | $\overline{dCu_1\text{-}Cu_1}$ (%) | $\overline{dN\text{-}N}$ (%) |
|---|---|---|---|---|---|---|---|---|
| 4 | 1 | $(5\sqrt{2}\times\sqrt{2})R45°$ | 0.131 | 0.085 | 0.115 | +6.7 | +2.3 | +4.4 |

As the N coverage increases towards saturation, the stripe width approaches monoatomic-thickness before eventually disappearing. Our stress model in Equation (18) indicates that the average compressive stress level in the thinnest striped surface, at the experimental N-patch length scale, is substantial (~ 3 N/m). Recently, Yamada *et al.* proposed the possibility of a missing-row formation in a narrow stripe boundary.[3] Such a missing-row boundary is different from the trench that forms approaching saturation coverage in that N atoms are not present in the boundary. In addition, the missing row boundary is aligned along the <100> direction. Our model for the missing-row boundary is shown in Fig. 4. The surface unit cell used for the calculation of the missing-row boundary is of the same dimension as that for the striped structure in Fig. 2d (*l=4, d=1*). We present our calculated surface stress levels in Table 7 as well as the relaxation parameters in Table 8. The stress relief achieved by the missing-row boundary is remarkable, even the absolute stress level turning from compressive to tensile. As compared with the simple stripe boundary of a clean Cu row (*l=4, d=1*) in Table 6, the stress reduction via the missing-row boundary is by far larger. As expected, this large stress relief results from the large lateral Cu-Cu and N-N expansions owing to the holes created in the boundary together with the absence of N atoms in the proximity.

The question thus arises: what are the energetic costs for creation of the (simple) stripe and missing-row boundaries and thus which is preferred. Hence we calculate the boundary creation energy per area for both. For simple stripe, the boundary creation energy per area ($E_{bc}$) is calculated as:

$$E_{bc} = \frac{1}{A}(E[\text{ideal c}(2\times 2)\text{ N/Cu(001)}] + E[\text{clean Cu(001)}]$$
$$- E[\text{c}(2\times 2)\text{ N/Cu(001) with stripes}]$$
$$- E[\text{a row of N on clean Cu(001)}]) \tag{19}$$

where E[system] represents the total energy of the corresponding system and A is the surface area of a used supercell. We used the same surface unit cell to calculate the total energy of each system in (19), that is, *(5√2×√2)R45°*. Thus calculated boundary creation energy is equal to the work needed to create a strip boundary and therefore should be positive. Similarly, the boundary creation energy per area is calculated for missing-row boundary, as:

$$E_{bc} = \frac{1}{A}(E[\text{ideal } c(2\times2) \text{ N/Cu(001)}] + 2*E[\text{clean Cu(001)}]$$
$$- E[c(2\times2) \text{ N/Cu(001) with missing-row boundary}]$$
$$- E[\text{a row of Cu on clean Cu(001)}]$$
$$- E[\text{a row of N on clean Cu(001)}]) \quad (20)$$

Our calculated $E_{bc}$ for stripe-boundary and $E_{bc}$ for missing-row-boundary are 1.6 and 16.4 meV/Å$^2$, respectively. Thus, the creation of stripe boundary of clean Cu row is by far easier than the creation of missing-row boundary. The inference is that at low coverages stripes should form in the boundaries. However, as N coverage increases (and, with it, N-induced elastic repulsive interaction), the chance of the formation of a missing-row boundary must grow as well. Importantly, the surface phase with missing-row boundary can be considered a transition phase from striped surfaces to surfaces with the formation of trenches at saturation coverage. We will discuss this point in detail below.

### 5. Surface stresses of the *c(2×2)* N/Cu(001) surfaces with trenches at saturation coverage

Table 9. Calculated surface stress for surfaces with trenches.

| Surface phase | Unit cell | $l$ ($a_o$) | $d$ ($a_o$) | Analytical stress (N/m) | Numerical stress (N/m) |
|---|---|---|---|---|---|
| trenches only: along ⟨100⟩ | (2√2×√2)R45° | 1 | 1 | -0.40 | -0.23 |
|  | (5√2×√2)R45° | 4 | 1 | -1.58 |  |
| trenches only: along ⟨110⟩ | p(2×2) | 1/√2 | 1/√2 | +0.23 | +0.32 |
|  | p(4×2) | 3/√2 | 1/√2 | +0.10 |  |
|  | p(6×2) | 5/√2 | 1/√2 | -0.42 |  |

Table 10. Theoretical structural parameters of c(2×2) N/Cu(001) surfaces with trenches along <110> direction for different trench width d and $l=L-d$.

| $l$ ($a_o$) | $d$ ($a_o$) | unit cell | $\overline{dN\text{-}Cu_1}$ (Å) | $\overline{r_1}$ (Å) | $\overline{\delta}$ (Å) | $\overline{\Delta_{12}}$ (%) | $\overline{dCu_1\text{-}Cu_1}$ (%) | $\overline{dN\text{-}N}$ (%) |
|---|---|---|---|---|---|---|---|---|
| $1/\sqrt{2}$ | 1 | p(2×2) | -0.111 | 0.0 | 0.0 | +4.0 | N/A | N/A |
| $3/\sqrt{2}$ | 1 | p(4×2) | 0.018 | 0.106 | 0.083 | +6.9 | +4.9 | +2.1 |
| $5/\sqrt{2}$ | 1 | p(6×2) | 0.058 | 0.043 | 0.116 | +8.9 | +3.8 | +1.8 |

Table 11. Theoretical structural parameters of $c(2\times2)$ N/Cu(001) surfaces with trenches along <100> direction for different trench width d and $l=L-d$.

| $l\,(a_o)$ | $d\,(a_o)$ | unit cell | $\overline{dN\text{-}Cu_I}\,(\text{Å})$ | $\overline{r_1}\,(\text{Å})$ | $\overline{\delta}\,(\text{Å})$ | $\overline{A_{12}}\,(\%)$ | $\overline{dCu_I\text{-}Cu_J}\,(\%)$ | $\overline{dN\text{-}N}\,(\%)$ |
|---|---|---|---|---|---|---|---|---|
| 1 | 1 | $(2\sqrt{2}\times\sqrt{2})R45°$ | 0.069 | 0.007 | 0.146 | +3.9 | +6.1 | +2.4 |
| 4 | 1 | $(5\sqrt{2}\times\sqrt{2})R45°$ | 0.130 | 0.042 | 0.114 | +6.9 | +2.1 | +2.1 |

FIG. 10. Calculated stress levels for several c(2×2) N/Cu(001) surface phases with respect to the distances between trenches $l$. The surfaces considered in this graph have trench width d, of $1/\sqrt{2}\,a_o$ only. For details refer to Table 9.

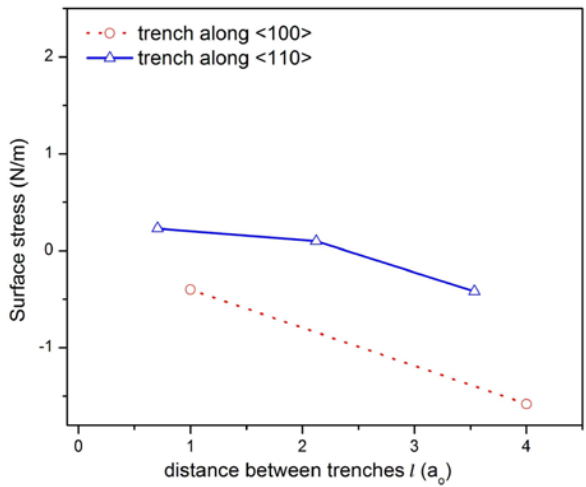

At saturation N coverage, the stress mechanisms so far discussed become unavailable. Therefore, some completely new mechanism is required for stress relief. In contrast to stripes, trench formation modifies Cu density by working a defect (hole) into the substrate superstructure. The N-coverages can thus remain the same, that is, 0.5 ML. To examine the effect of trench direction on stress changes, we model trenches, both as observed in experiment along the ⟨110⟩ direction, as in Fig. 5a-c and along the ⟨110⟩ direction, as in Fig. 5d-e. For each case we create a spatial periodicity of trenches by removing a row of Cu atoms along the direction in question and calculate the surface stresses. We have reported our calculated stress levels in Table 9 and Fig. 10 as well as structural parameters in Table 10 for surfaces with trenches along the <110> direction and in Table 11 for surfaces with trenches along the <100> direction. By comparing two curves in Fig. 10, it is clear that a trench along the <110> direction (triangles in Fig. 10) is much more effective in relieving stress than a trench along the <100> direction (circles in Fig. 10). In contrast to what happens with stripes (See Fig. 9), as $l$ approaches zero, the stress level of the trench surface does not approach that within a clean surface, but rather stays closer to zero. This happens because the Cu vacancy constituting the trench works to relax mainly compressive stress since expanding Cu-Cu bond lengths can easily adjust, through the trench, to a compressive stress within the substrate.

It might seem that a comparable effect could be claimed for the trench along the <100> direction, which also creates a Cu vacancy. However, the distance between the Cu rows across the trench in the latter is $\sqrt{2}$ times smaller (cf. Fig. 5a) than that created by the trench in the former direction (cf. Fig. 5d). In other words, there is more space for relaxation of the $Cu_1$-$Cu_1$ lateral bond in the former. As a result, the lateral $Cu_1$-$Cu_1$ distance $d(Cu_1$-$Cu_1)$ for the trench along <110> (for example, 3.8% of the bulk lattice constant for $l=5/\sqrt{2}\ a_o$, $d=1/\sqrt{2}\ a_o$ in Table 10) is larger than that for the trench along the <100> direction (for example, only 2.1% of the bulk lattice constant for $l = 4\ a_o$, $d=1\ a_o$ in Table 11). Moreover, Cu atoms are more densely packed per unit length along the <110> direction than any other direction including the <100> direction. Thus, removing Cu atoms along the <110> direction is the most effective way to reduce Cu density at the surface. All these facts help explain why trenches along <110> directions are more effective than trenches along <100> directions at relieving stress. (Note that the same arguments can be equally applied to stripe formation, in which case the <100> direction is favored over any other direction.)

To compare the energetics, we calculated the trench-creation energy along lines similar to what we did in the case of stripes, in (19) - (20). The surface unit cells used are $p(2\times6)$ for trenches along the <110> direction and $(5\sqrt{2}\times\sqrt{2})R45°$ for trenches along the <110> direction. While the trench densities are comparable, we find trench-creation energies are 17.0, and 29.0 meV/Å$^2$ for the <110> and <100> directions, respectively. Thus, at this trench density the trench along the <110> is favorable not only in terms of stress reduction but also in terms of energetics. While both of the calculated trench-creation energies are far larger than that for stripe-creation (1.6 meV/Å$^2$), the creation energy of a trench along the <110> is comparable to or slightly larger than that of the missing-row boundary along the <100> direction. Thus, the preference order of the boundary creation energy is stripe > missing-row > trench along the <110> direction > trench along the <100> direction. This order, in fact, may reveal the order in the phase transition at the c(2×2) N/Cu(001) surface: While at low N coverage surface stress within the N patches is relieved by stripes as N coverage increases up to saturation, since the space for stripe formation is increasingly limited, stress relief by stripe formation eventually becomes ineffective, so that a new stress relief mechanism appears: the missing-row boundary forms in the narrow monoatomic thick stripe region, thereby enabling further stress reduction within the N patches. Finally at saturation coverage, trench nucleates along the <110> direction. Note that a trench along the <100> direction may not form a c(2×2) N/Cu(001) surface since trenches along the <110> direction already effectively relieve compressive stress levels at saturation coverage. In this picture, the missing-row boundary is an intermediate phase between stripe phase at subsaturation coverage and trench phase at saturation coverage. Since a missing-row boundary and a trench are in essence one (i.e. a Cu vacancy), this phase can certainly be considered as a concurrent phase of stripes and trenches.

In fact, this picture is well supported by our HAS experiment, in which the averaged lattice parameter within N-patches initially reduces, with increasing N coverage, from 3.5% to 1.8% above that of the substrate lattice parameter, indicating that stress relief is increasingly constrained with decreasing width of stripes, as occurs as N coverage increases. The N-containing patches, however, are not compressed beyond a certain 1.8% expansion level. We attribute this critical contraction level to the onset of missing-row formation followed by trench nucleation at saturation.

In summary, trench and stripe formation stress-relief mechanisms are quite effective. It is clear, therefore, that the concurrent formation of stripes and missing-Cu rows (or trenches) at high N coverages is expected to be by far more effective in relieving surface stress. The coexistence of stripes and missing-Cu rows can therefore serve to maintain low absolute surface stress levels. As confirmed with our He diffraction findings, with increasing N coverage (above 0.35ML) the averaged N-lattice parameter would be expected to rise, because the trenches can enable proportionately larger patch relaxations. The *c(2×2)*-C/Ni(100) surface exhibits a similar trend – i.e., zero stress change for C coverages from 0.34 ML to 0.43 ML, during which Ni substrate atoms undergo clock reconstruction.[44]

## IV. CONCLUSION

We have calculated the surface stress of *c(2×2)*-N overlayers on Cu(001) using density functional theory within the pseudopotential approximation. Upon N adsorption, substrate Cu atoms in the outermost layer do not undergo as significant vertical displacements (rumpling) as proposed in an earlier study, which offered such a model to account for stress relief in this system.[45] An optimized N-induced *c(2×2)* structure has a net surface stress level of ~ 4 Nm$^{-1}$. Our calculations demonstrate that rumpling displacements within the outermost Cu layer do not act to relieve the compressive surface stress. And though clock displacements *could* relieve lateral stress levels substantially, we find that they are not energetically viable. We find instead that, although such stress is somewhat relieved when trenches of missing Cu atoms form along the <100> direction, it is most effectively relieved when stripes of clean Cu(001) form along the ⟨100⟩ direction or when trenches of missing Cu atoms form along the ⟨110⟩ direction.

He diffraction experiments have shown that the surface strain within N-containing patches initially is reduced with increasing patch density. Calculations of stress levels within 1-D models for alternating patch and striped structures have indicated that N-containing regions are less compressible than N-free regions. This deduction appears also to be borne out qualitatively by others' images of N-containing patch distributions on the Cu(001) surface.

He diffraction also indicates that the N-patch strain levels increase as stresses are further relieved with inclusion of the missing Cu row trenches in the surface. We have concluded that the coexistence of stripes and trenches serves to limit average surface stress levels, by enabling larger lateral relaxations of Cu atoms within the uppermost surface plane.

We hope that our work will motive some more experiments that could directly measure the stresses and/or confirm our reconstruction model. In particular a cantilever based atomic force microscope measurement could help validate some of the predictions in this work.

**ACKNOWLEDGEMENTS**

This work was supported in part by NSF grant CHE 0718055 and by DOE grant DOE Grant DE-FG02-07ER46354. The authors would also like to thank David Vanderbilt for discussions on spatial distributions of regions with imbalanced surface-stresses. We are grateful to Lyman Baker for careful reading of the manuscript and many constructive comments.